# Beyond networks, towards adaptive systems

Luiz Pessoa
Department of Psychology, University of Maryland, College Park, MD 20472, USA
pessoa@umd.edu
ORCID iD: 0000-0003-4696-6903

**Abstract**

Despite their widespread utility across domains, basic network models face fundamental limitations when applied to complex biological systems, particularly in neuroscience. This paper critically examines these limitations and explores potential extensions and alternative frameworks better suited to capture the adaptive nature of biological systems. Key challenges include: the need to account for time-varying connections and adaptive topologies; the difficulty in representing multilevel systems with cross-level interactions; the inadequacy of fixed state spaces for systems that continuously expand their range of possible states; and the challenge of modeling deep history dependence and open-world interactions. While some limitations can be partially addressed through extensions like multilayer networks and time-varying connections, I argue that biological systems exhibit radical context dependence and open-endedness that resist conventional mathematical formalization. Using examples primarily from neuroscience, I demonstrate how even sophisticated approaches incorporating dynamics and context sensitivity fall short in fully capturing biological complexity. Moving forward requires developing frameworks that explicitly account for adaptive reconfiguration, embrace open-ended state spaces, and draw inspiration from concepts like Kauffman's "adjacent possible." The path ahead demands mathematical and computational approaches that are fundamentally more dynamic and flexible than traditional network models.

# 1 Introduction

Network models have been remarkably fruitful in modeling a broad range of phenomena across biological, social, and technological domains (Newman, 2018). In biology, they have elucidated gene regulatory networks and ecological food webs. Social network analysis has transformed our understanding of human interactions, studying phenomena like information spread and community formation. In technology, network models have been applied to computer networks and the structure of the Internet.

However, despite their widespread utility, I argue that basic network models face significant limitations when applied to *complex biological systems*. This paper aims to critically examine these limitations and explore potential extensions and alternative frameworks that may better capture the properties of adaptive biological systems. In the ensuing presentation, I will focus mostly on examples from neuroscience.

To ground our discussion, let us first define what we mean by a "basic network". Consider a standard network defined in terms of a set of nodes and a set of edges (Newman, 2018). The nodes can represent any entities of interest: genes, persons, computers, etc. Edges link nodes to one another and formalize their relationships, such as linking genes in common molecular pathways, persons who know each other, or computers in a local network. Network models have galvanized the study of collective properties of a broad array of systems, thus uncovering organizational principles previously unknown based on the study of their subcomponents.

While this basic formalism is powerful in its generality, it faces several key challenges when applied to complex biological systems. The limitations of basic network models discussed in this paper include:

1. The need to account for time-varying connections and adaptive topologies.
2. The need to model variables and their dynamics.
3. The challenge of representing multilevel systems with cross-level interactions.
4. The difficulty in capturing the radical context dependence exhibited by many biological systems.
5. The inadequacy of fixed state spaces for systems that continuously expand their range of possible states.
6. The challenge to represent the deep history dependence often observed in biological phenomena.
7. The challenge of modeling open-world interactions as opposed to closed worlds.

This paper will examine each of these challenges, discussing how they manifest in biological systems, particularly in neuroscience. We will explore various attempts to extend network models to address these issues, including time-varying connections, multilayer networks, and adaptive topologies.

For each limitation, we will consider how possible extensions of network formalisms succeed or fail in addressing the challenge. In addition, we will explore alternative modeling approaches that may better capture the adaptive nature of biological systems. These include dynamical systems frameworks, data-driven techniques for learning system dynamics, and concepts from complexity science like Kauffman's "adjacent possible".

It is important to note that the field of network science has developed numerous sophisticated techniques beyond "basic networks". Thus, the extent to which the critiques in this paper apply to "network science" as a whole depends on how broadly one defines the field. Regardless of terminology, many of the considerations discussed here apply to mathematical formalization of biological systems in general.

By critically examining the limitations of current modeling approaches, this paper aims to stimulate the development of new frameworks that can better capture the complexity, context-sensitivity, and open-ended nature of biological systems. While I do not advocate abandoning network models entirely, I argue for pushing them in more adaptive and flexible directions. Ultimately, advancing our understanding of complex biological phenomena, including brain function and behavior, requires mathematical and computational frameworks that are radically more dynamic and open-ended than traditional network approaches.

In the next section, I briefly define and discuss basic network models. In the subsequent sections, I gradually introduce changes to the basic scheme so as to highlight how these features are necessary to model biological systems. This gradual buildup not only provides a review of how basic network models can and have been expanded, but also describes when the necessary extensions hit roadblocks that challenge the effectiveness of using network models when applied to dynamic, multi-level, contextually dependent, and open-ended biological systems.

## 2 Some challenges to network models

A *basic network* is described in terms of a graph $G = (V, E)$, which is defined in terms of a set of nodes (or vertices) $V$ and a set of connections (or edges) $E$. The set $V$ specifies all elements of interest (e.g., brain regions) and the edges are often described in terms of an adjacency matrix $A$, whose entries $a_{ij}$ represent the connection between nodes $i$ and $j$ (e.g., regions $i$ and $j$ are anatomically interconnected). These connections can be binary (0, no connection; 1, a connection exists) or vary continuously in strength.

In the context of brain research, basic networks are most often employed in two contexts, namely, anatomical and functional. In both cases, the networks consist of a) a fixed set of nodes, typically brain regions $R_1, \ldots, R_n$, and b) a fixed set of pairwise relationships between nodes, such as the presence/absence of an anatomical connection between regions $i$ and $j$ in the case of anatomical networks, or a functional relationship between these regions, such as the extent to which their signals are correlated across time.

In network models, the nodes define them most *basic units of interest*. Nodes have no internal properties; what differentiates nodes $p$ and $q$ is their set of edges. Thus, in essence, nodes "don't do anything", except indicate the existence of an element, which is odd because, for example, a key goal of subdividing the brain in terms of anatomical regions (also called areas) is the fact that they are thought to do different things. As stated, the framework dictates that nodes, in some fashion, encapsulate the basic units (atoms) of interest. In other words, the level of granularity of the nodes (e.g., regions) is the rightful one to explain the phenomena of interest. Thus, if a node is not included, it is by definition excluded from the set of entities that have an "elementary" explanatory role. In the case of the brain, it amounts to saying that, say, region $p$ does not participate in the processes of interest. Likewise, network edges define the most *basic relational* units of interest. For example, the anatomical or functional relationships between brain areas. The implication is that by utilizing the relationships between nodes, system-wide properties can be obtained based on this type of linkage information.

As in other areas of research, application of basic network models has provided many insights into large-scale brain organization (Sporns, 2016; Fornito et al., 2018). For example, network analysis has revealed that certain brain regions act as "hubs," having a high number of anatomical connections to other regions. These hubs tend to be more densely interconnected with each other than expected by chance, forming a "rich club" organization. This rich club structure has been found to play a crucial role in global brain communication and integration of information across different functional domains.

## 2.1 Time-varying connections

Anatomical brain networks are relatively static, as the presence of a fiber bundle linking two areas is fixed in the adult brain, and its strength changes on a slow time scale. Thus, it's justifiable to analyze a single graph when studying anatomical connections. But of course, in many biological systems, the relationships between nodes evolve over time. For example, the functional relationship between two neurons in a local circuit can change on a fast time scale. Thus, the first modification to the basic network scheme that we will consider here is to consider the adjacency matrix to be time varying, $A(t)$. If the connections between nodes $i$ and $j$ admit strengths that vary continuously, we can denote them as $w_{ij}(t)$.

Although the introduction of time-varying connections is conceptually simple, it obviously expands the range of phenomena that can be studied with network models. By having multiple graphs, $G_t = (V, E_t)$, one for each time $t$, it is possible to simply repeat the analyses across graphs and identify network characteristics that change across time (e.g., in newborn and adult brains). More generally, this is an active area of research, and multilayer networks can be used to capture different temporal snapshots that are analyzed simultaneously (Boccaletti et al., 2014). This allows the analysis of *network temporal evolution*, such as studying changes in network organization over time, including identifying the emergence or disappearance of cohesive clusters of nodes called communities[1], and analyzing temporal patterns of node importance (properties such as centrality).

## 2.2 Dynamics at the level of units: nodes with variables

The vast majority of network models have objects, notably nodes, that are devoid of content. That is to say, every node does the same thing. In fact, they are mere "place holders" and in this way are the simplest form of mathematical variable possible, so-called "indicator variables" (i.e., they indicate that an object exists). The implication is that network models only consider collections of homogeneous objects. This is a severely limiting assumption, especially in the case of biological systems which have rich structural organization and diversity.

---

[1] For example, our studies of functional networks based on functional MRI data have described how experimental conditions involving the threat of shock lead to changes in network-level functional organization (McMenamin et al., 2014).

A natural way to extend the representational power of basic network models is to consider that every node represents a variable (or variables) of interest that evolves according to an associated *dynamics equation*. For example, if object (node) $p$ does something different from object (node) $q$, this can be represented in terms of the variables $x_p$ and $x_q$ and the fact that the nodes perform different functions $F_p$ and $F_q$. Mathematically, one can consider how objects evolve temporally by specifying their dynamics in terms of differential equations:

$$\frac{dx_p}{dt} = F_p(x_p).$$

For example, the equation might indicate that the variable of interest undergoes exponential decay (or growth); in the case of the brain, the equation could represent how the activity of region $p$ evolves in time. If the reader is not familiar with descriptions in terms of differential equations, one can consider a discrete-time version instead: $x_p(t + 1) = F_p(x_p(t))$, that is, the future value at node $p$ depends on the current value as transformed by the function $F_p$.

In network models, as nodes are interconnected, the temporal evolution of the variable at a given node depends on the joint values across multiple variables (nodes). Specifically, the dynamics of a node $i$ depends on the dynamics of all nodes $j$ to which it is connected. For example, the population of a prey species and its predators mutually influence each other, such that, say, (all else equal) the greater the number of predators, the fewer the prey; and the fewer the predators, the larger the number of prey. We can state this type of relationship as follows:

$$x_p(t + 1) = F_p\left(x_p(t)\right) + \sum_g F_g\left(x_g(t)\right).$$

In other words, the temporal evolution of node $p$ depends on its previous state and the sum of the contributions of the previous states of all nodes $g$ to which it is connected (other nonlinear interaction schemes are possible, of course, but the additive description suffices for illustration).

Naturally, the extension to basic network models described in this section amounts to using mathematical models of coupled differential (or difference) equations, which constitute the bread-and-butter of formalisms used in physics and applied mathematics. In the context of network models, however, the considerable gain in expressivity comes at a high cost. If every object of the system evolves according to its own dynamics, many of the techniques of network science no longer apply (e.g., community detection algorithms, graph measures such as centrality, etc.).

The argument I want to make in this section is that traditional network models, if they are to be successfully applied to the study of biological systems, need to be extended so as to describe multiple interlinked variables simultaneously. To do so requires nodes to represent specific variables and to specify their dynamics. In network science, work in this area is very limited, and further developments are needed. A recent attempt to address this gap was outlined by Artime and De Domenico (2021), where each node in a network was characterized by a set of features; for example, whether a node was "infected" or "susceptible" in the context of epidemiological modeling[2].

## 2.3 Multilevel systems

Now, let's turn to a considerably greater challenge to network models, namely, capturing properties of systems spanning multiple levels[3]. Network models frequently focus on a specific domain of interest, such as genes, brain regions, or computers. However, biological systems are inherently multilevel, as vividly illustrated in textbooks. A neuroscience textbook will typically discuss the contributions of genes, molecules, cells, local cell circuits, brain areas, and entire brain sectors (e.g., prefrontal cortex). Naturally, not all levels are considered when discussing a given behavior or mental state, and specializations within the field reflect this focus. Consider hunger. Cognitive neuroscientists might be interested in brain areas that play notable roles in such state/behavior, such as the hypothalamus and specific cortical areas. Systems neuroscientists will focus on circuits within the hypothalamus, or possibly across multiple areas. Molecular neuroscientists will focus on molecules that play pivotal roles in hunger and feeding, such as ghrelin and leptin. And so on.

Explanations in neuroscience frequently focus on a single level, or possibly a pair of adjacent levels, such as molecules (e.g., neurotransmitters) and cells (see below). Scientists often target particular levels for pragmatic reasons, primarily based on the measurement technique they employ. For example, tungsten electrodes can be inserted into brain tissue to measure electrical properties of neurons, or a carbon fiber microelectrode can be inserted into a region to estimate the concentration of a molecule such as dopamine. Scientists who use neuroimaging techniques such as functional magnetic resonance imaging (fMRI), focus on mesoscale phenomena at the brain region/subregion level given that the technique typically measures signals in a brain volume of (currently) around 10 mm$^3$ (N.B.: a single fMRI "pixel" contains hundreds of thousands of neurons).

Another important reason researchers focus on particular levels of explanation is related to their putative *explanatory primacy*. Because neuroscientists have historically considered the neuron to be the key explanatory unit, they have naturally sought to measure signals from these cells. Neuroscientists have also viewed brain regions as functional units. Accordingly, the use of fMRI to study responses that can be linked to such units has played a significant role in studying the brain.

Nevertheless, in neuroscience, and indeed biological systems, deeper understanding requires consideration of processes at multiple spatiotemporal scales, including subcellular, cellular, and multicellular levels. In practice, not all levels can be considered simultaneously, not only due to practical constraints but because not all levels may be relevant to a given research question. Thus, in practice when the research goal calls for focusing on a given level, $L_i$, surrounding levels are typically considered: …, $L_{i-1}$, $L_i$, $L_{i+1}$, …. In neuroscience, when focusing on the neuronal level, it is important to consider the lower level of molecules (e.g., neurotransmitters) and the higher level of local circuits. Consider the dopamine system. Dopamine is synthesized in cells in the ventral tegmental area in the midbrain that project to the nucleus accumbens in the ventral striatum. Neurons in the latter have receptors that are sensitive to dopamine molecules. But there are different dopamine receptor types in the ventral striatum, and these can have different impact on cell activity. In particular, cells with D1 receptors will often be excited by incoming dopamine release and tend to fire more vigorously. But more generally the effects are complex and depend on the type of receiving neuron (determined by their receptor type called D1, D2, and so on) and their local interconnections. Thus, explanations require investigation of circuit-level properties, too.

---

[2] Whether mathematical models that are purely based on sets of coupled differential (or difference) equations count as "network models" is a question of semantic and terminological preference. I find this to be a stretch but others might legitimately disagree.
[3] The issue of levels is rather complex, as extensively discussed in the literature. Although it will not be possible to discuss this topic at greater length here, as should be clear from the discussion in this section, I do not consider "levels" as insulated from one another (Wimsatt, 2007; Brooks et al., 2021). In fact, part of the challenge faced by network models lies in the difficulty of capturing such cross-level interdependence.

The preceding example illustrates how understanding in neuroscience invariably requires considering multiple levels of analysis (Figure 1). The implication for network models is that they would benefit from adopting multilevel approaches. Indeed, multilevel network models have received attention in sociology (Zappa and Lomi, 2015; Chabert-Liddell et al., 2021) and ecology (Pilosof et al., 2017). In neuroscience, despite calls for more multilevel approaches (Presigny and Fallani, 2022; Shine et al., 2021; Grosse-Wentrup et al., 2023), multilevel network models are relatively rare (see also Castiglione et al., 2014; Aftab and Stein, 2022; Liao et al., 2024).

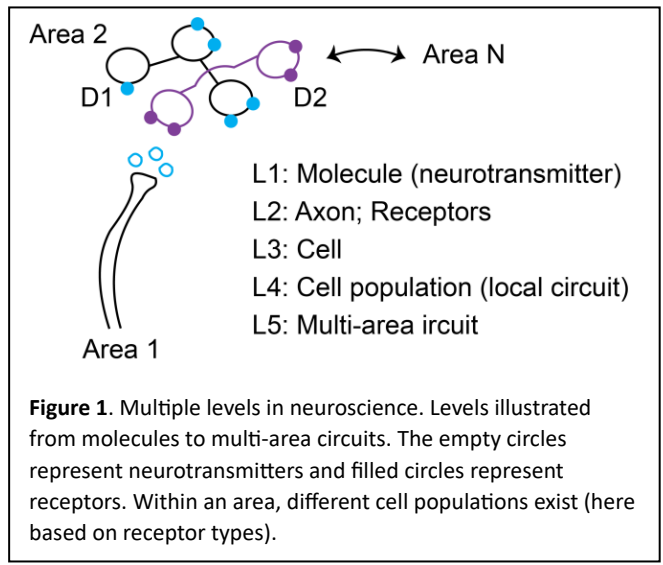

**Figure 1**. Multiple levels in neuroscience. Levels illustrated from molecules to multi-area circuits. The empty circles represent neurotransmitters and filled circles represent receptors. Within an area, different cell populations exist (here based on receptor types).

A further consideration is that, in biological systems, not only are multilevel models needed, but they should capture interlevel interactions from lower to higher levels (as per usual) *and* from higher to lower levels (more controversially). Noble (2011) discusses examples of the latter, including higher levels of organ and tissue controlling gene expression, as well as triggering cell signaling. Although these diverse forms of *downward influence* are at times controversial given debates about the ontological status of downward causation, here I propose to take a pragmatic approach instead[4]. Consider a fundamental result in neuroscience, that of the model of neuronal firing by Hodgkin and Huxley (1952). In the model, equations are used to describe the behavior of cell-level membrane potentials in terms of the opening and closing of ion channels (molecular-level entities). The membrane potential, a cell-level property, directly influences the opening and closing of ion channels, which are molecular-level entities. Conversely, the state of these channels determines the ionic currents, which in turn alter the membrane potential. This creates a feedback loop where the cell-level voltage and the molecular-level channel states continuously influence each other, resulting in the phenomenon of the action potential. Neither level can be said to be privileged; rather, it's their interaction that produces the observed behavior.

---

[4] In other words, here we can focus on the epistemic utility of viewing higher and lower level as interdependent for the purposes of explanation, including prediction. My point here is simply to avoid issues of "downward causation" and "emergence" that, important as they are, need not be addressed here.

Taken together, I argue that biological systems need explanatory frameworks that *simultaneously* consider multiple levels of organization, as others have argued in the past (Kitano, 2002; Castiglione et al., 2014; Noble, 2011).

## 2.3 Context dependence and adaptive topologies

A key property of basic networks is that they are comprised of a fixed set of entities (nodes), as captured by its mathematical definition, $G = (V, E)$. While this appears to be an innocuous property, consider what having fixed nodes entails: one is able to enumerate all elements that matter in a given situation *in advance* (e.g., the set of all brain areas). Put another way, the atoms of interest can be defined and provide the necessary parts-list required to specify the model. Whereas this may be feasible in some cases, in general it is not possible to list all the parts in advance in biological systems.

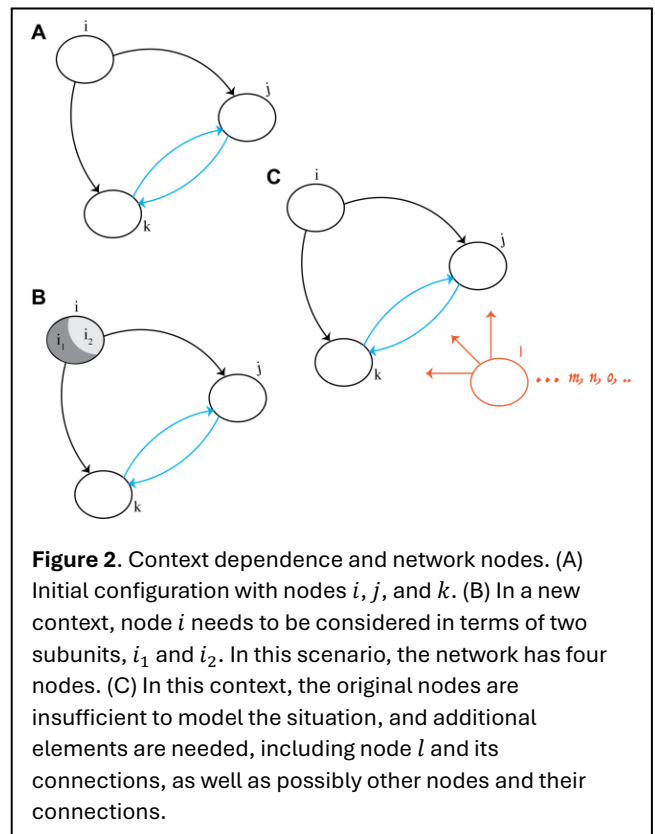

To illustrate, consider a network with nodes $\{i, j, k\}$ (Figure 2) (a small network is used for illustration but naturally one deals with very large networks in many practical problems). While the network may specify the elements of interest in some contexts, suppose that in a new situation two subcomponents of node $i$ must be considered: in the new context, one needs to consider subunits as the proper units of interest (thus the nodes of interest would be $\{i_1, i_2, j, k\}$). In yet another context, in addition to the original elements, perhaps an additional node needs to be considered (so the nodes would be $\{i, j, k, l\}$). We will further discuss this challenge below, but for the time being let's consider this to be a problem of *unanticipated context effects*.

**Figure 2**. Context dependence and network nodes. (A) Initial configuration with nodes $i$, $j$, and $k$. (B) In a new context, node $i$ needs to be considered in terms of two subunits, $i_1$ and $i_2$. In this scenario, the network has four nodes. (C) In this context, the original nodes are insufficient to model the situation, and additional elements are needed, including node $l$ and its connections, as well as possibly other nodes and their connections.

The reader might object that this is not a substantive problem. After all, the unanticipated contexts simply point to the state of ignorance of the modeler, who should have included all the necessary elements at the outset. Although this appears to be an effective defense, it requires the modeler to have a *complete parts-list* for the problem of interest. I argue that this is not feasible for biological problems in general, because of what I term the property of *radical context dependence* of biological systems, a theme further developed below[5].

As a more concrete example, consider a network model at the level of brain regions, a type of decomposition that may be reasonable in particular instances. Yet, the definition of brain regions has evolved historically, and is far from a settled question (e.g., Hayden, 2022). In particular, separate neuronal populations within a brain area may need to be considered. For example, some neurons may have receptors for one of the main neurotransmitters (e.g., dopamine) but other neurons may have receptors for a different neurotransmitter (e.g., serotonin). Thus, projections to the area that release dopamine and/or serotonin will generate substantially different behavior *in the area*. Taken together, in some contexts the separate neuronal populations may be engaged in a manner that partitions the region into (at least) two functional subareas (which would then constitute the relevant units of interest).

As pointed out, the reader may object and point out that the area should have been considered to have two (or possibly more) subpopulations at the outset. The response to this assertion is that brain areas have a very large number of cell populations, and one does not know *in advance* when they need to be considered. Of relevance here, recent studies have uncovered >3,000 types of neurons in the brain (Siletti et al., 2023). At the very least, this issue leads to an *enumeration explosion*, where the investigator faces the challenge of having to consider all conceivable sources of influence. It should be pointed out that cell-type diversity in brain areas simply provides an example of how context-dependent processing is instantiated in the brain. It provides a stark example of the challenges for network models of brain function.

More generally, I argue that biological systems exhibit considerable---and potentially radical---context dependence, as further illustrated in the next section. Thus, in practice, although network units of interest (i.e., nodes) can be outlined for a few contexts, $C_1, ..., C_n$, I argue that a better approach is to explicitly consider the elements to *vary as a function of context*. In the graph notation above, we can denote this as $G_{c,t} = (V_c, E_t)$, where $c$ is context and $t$ is time. What is needed, therefore, is for investigators to employ adaptive network topologies (i.e., that grow and/or shrink), in fact that can change dynamically as a function of context. Theoretical work on adaptive network topologies is in its infancy, but some proposals have been outlined; see Gross and Sayama (2009), in particular the work by Sayama and colleagues (e.g., Chapters 14 and 15 of this reference).

## 2.4 Context dependence in the brain

Let's take this section to further illustrate the striking context sensitivity of function found in biological systems by considering a few examples from neuroscience. Consider some of the ways chemicals known as neuromodulators impact circuit function and brain/behavioral states. These substances include serotonin, norepinephrine, dopamine, acetylcholine, and histamine, a large number of peptides, as well as hormones, endocannabinoids, among others. The field of ethology has studied behavioral states for many decades, notably those related to reproductive behavior, foraging, feeding, aggression, defensive behaviors, exploration and exploitation, social behaviors, navigation, and play behaviors. Neuroscientists have investigated the neural circuits and mechanisms that support such classes of behavior, and uncovered the central roles exerted by neuromodulators.

---

[5] It could be agued that some mathematical formalisms make the need for networks that would add/delete nodes unnecessary. We will discuss some of these models with latent states, including probabilistic models such as Hidden Markov Models below.

In both biological and artificial recurrent neural networks, neuromodulatory inputs can rapidly and flexibly impact the conduct of a particular circuit so that it can instantiate and/or participate in different functions. In particular, neuromodulators can modify circuit processing by modulating synaptic gain, plasticity, and neural excitability, among other mechanisms (Dayan, 2012; Marder et al., 2014; Nadim and Bucher, 2014). In the context of the present paper, we can think of these effects as impacting a node's function/parameters (as in the case of excitability) and/or a connection's function/parameters (as in the cases of synaptic gain and plasticity). Importantly, neuromodulators have the ability to alter the function of an entire circuit; that is, local changes at the level of nodes and connections are capable of impacting collective (emergent) circuit functions. For example, in a computational model of the motor cortex, Stroud/Vogels et al. showed that simply changing neuronal gain (i.e., how neurons respond to inputs) enabled the exact same circuit (even with the same connection weights) to exhibit substantially different functions.

An example of the power of neuromodulation to alter and shift entire behavioral states is illustrated by a study that utilized whole-brain calcium imaging at single-cell resolution in unconstrained larval zebrafish to investigate the neural basis of foraging behavior (Marques et al., 2019). During hunting of live prey (Paramecia), zebrafish exhibited spontaneous alternation between two distinct and persistent states. The "exploitation state" was characterized by reduced locomotion and increased hunting behavior, resulting in spatially confined movements. In contrast, the "exploration state" involved enhanced mobility and diminished hunting, leading to expansive locomotion that facilitated exploration across space. The study identified a subpopulation of serotonergic neurons in the midbrain that displayed sustained activity and reliably predicted the exploitation state. These exploitation-state-encoding neurons were "triggered" by neurons in the forebrain, and the interaction between these two neuronal groups formed an oscillatory network that dictated the spontaneous alternation between exploration and exploitation states. Critically, the activity of this oscillatory network was associated with a comprehensive reconfiguration of sensorimotor processing during foraging, which resulted in pronounced alterations in both the drive to pursue prey and the precision of motor actions executed during hunting.

Examples of the impact of neuromodulation on circuit function like this one abound in the neuroscience literature (Brezina, 2010; Kehagia et al., 2010; Noudoost and Moore, 2011; Dayan, 2012; McCormick et al., 2020; Robson and Li, 2022). While developing this point further would take us too far field, the hope is that it provides a glimpse into the rich forms of context dependence in biological systems (for additional illustrations from my work, see also, Pessoa 2014, 2017, 2022).

## 3  Challenges to formalization more generally

Up to this point, I have considered how basic network models need to be extended to provide sufficiently powerful models of biological systems. This involved employing a) time-varying connections; b) incorporating dynamic variables; c) considering the inherent multilevel nature of biological phenomena; and d) dealing with the substantial context-dependence of biological systems. Different strands of network science have tackled these challenges to varying degrees of success[6]. In the following sections, I will focus on a set of issues that challenge the mathematical formalization of biological systems more generally. In other words, the problems discussed here pose difficulties not only to network models, but also to mathematical formalization in general (such as routinely used in physics and applied mathematics).

### 3.1 From fixed state spaces to more open-ended ones

To start developing the argument, it is useful to consider the concept of *state spaces*. In a nutshell, a state space is the "universe of the admissible", the set of states that the system in question can inhabit. Suppose a single particle exists within an enclosed box. The position of the particle can be described as $(x(t), y(t), z(t))$. The state space is the representation of the box in question in Euclidean space and contains all possible positions the particle can assume. In practice, the state space can be defined in terms of whatever variables of interest, which will correspond to coordinates in a typically high-dimensional space (e.g., a system with the populations of predators and preys along each axis). The concept of state spaces is powerful because the evolution of the system can be described as a *trajectory within the state space*. For example, the position of the particle may vary according to some law or equation that specifies its position at each time point, and the set of positions as a function of time constitutes the particle's trajectory.

Formal models described in physics and applied mathematics possess the critical feature that the state space is *fixed*: the variables of interest and their equations of evolution are immutable (i.e., laws), such that the "universe of the admissible" is formally specified. Let's break down the main implications of this statement:

1. Fixed state space: In traditional mathematical/physics models, once a state space is defined, it does not change. The dimensions (variables) are set and do not evolve.

2. Immutable variables: The variables that define the state space are predetermined and do not change throughout the analysis or evolution of the system.

3. Immutable equations: The laws or equations that describe how the system moves through the state space are also fixed. They do not adapt or change as the system evolves.

4. "Universe of the admissible": This universe refers to all possible states the system can occupy. In a fixed state space, this universe is completely defined at the outset and does not expand or contract.

---

[6] Although here it could be argued that one has to adopt a fairly permissive definition of "network science", one that includes a vast range of techniques from physics and mathematics.

5. Formal specification: All possible behaviors of the system are, in principle, knowable from the initial definition of the state space and its governing equations.

In brief, this perspective compels us to identify and understand all the relevant governing rules and variables that direct the behavior of the system under investigation *in advance*. Naturally, the success of physics attests to the power of this approach. But even in physics system dynamics can vary as a function of inputs or other variables. To account for such situations, system equations can be specified so as to vary with time: $f(\mathbf{x}, t)$. These systems are called non-autonomous. For the simplest example, consider a population that grows at a certain rate, $r$. If we know the population $N$ at time $t$, its growth is determined by $N(t + 1) = rN(t)$. Here the growth rate is fixed, but it can just as easily be specified to vary with time: we can use $r(t)$ instead of a fixed rate $r$ (e.g., greater than 1 during the breeding season but smaller than 1 during the non-breeding season).

In physics, a classic example of a non-autonomous system is the equation of motion for a forced damped pendulum. While the equation need not concern us here, the pendulum can exhibit a wide range of behaviors, including damped oscillations, forced oscillations, resonance, and chaotic behavior depending on how it is "forced" (i.e., perturbed).

While non-autonomous systems allow for time-varying dynamics that depend on inputs to the system, they still operate within a relatively fixed mathematical framework. In particular, the time-dependence of the equations must be specified *in advance*, and the possible ways in which the dynamics can change are limited by the precise form of the equations. This approach works well in classical physics, where only a few key conditions are outlined, given the focus on tractability and generalizability that is common in the discipline. It should be noted that the study of complex systems and computational physics require much larger numbers of "rules" and/or parameters. For example, in climate models, dozens of equations are employed, each with potentially multiple time-varying terms. The "rules" in this context are often embodied in how these terms are defined and how they interact, rather than being explicitly enumerated as in simpler non-autonomous systems.

To summarize, formal mathematical models applied to systems are most successful when the phenomenon in question is not strongly context dependent or involve a large number of interlinked levels, in addition to not exhibiting rich time-varying dynamics. However, biological systems are very frequently characterized by those very properties. In such systems, I argue that it is profitable to conceptualize the state space, itself, as *evolving dynamically*, a property that poses major challenges to existing mathematical formalisms. I call this property "open-endedness", as further elaborated next.

### 3.2 Open-endedness of biological systems

Many biological systems are much more open-ended than systems studied in physics. Their boundaries are not as clearly defined; they are less isolable---or, it could be said, they resist isolation. The property of open-endedness of biological systems goes hand-in-hand with their fundamental context- and history-dependence. Here, Kauffman's (2000, 2014) notion of the *adjacent possible* is a valuable framework to conceptualize the types of behavior exhibited by complex biological systems. His proposal was intended to apply to complex adaptive systems, particularly in the context of biological evolution and technological innovation. The core idea is that the range of possible future states or innovations in a system is constrained by its current state. The notion of the *adjacent possible* helps describe the set of all possible *next steps* that are available to a system at any given moment in time.

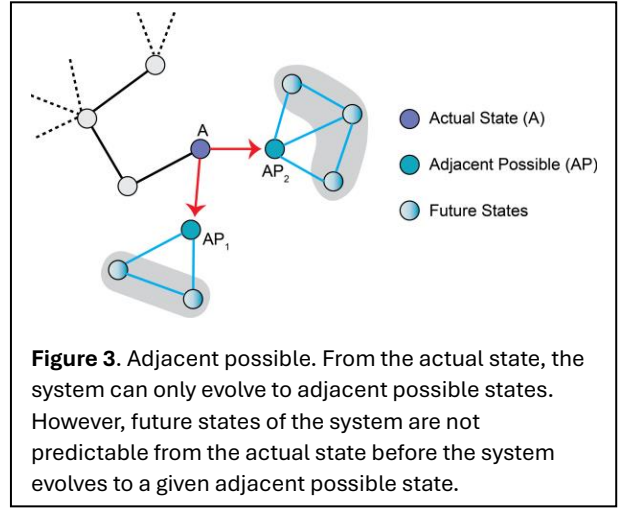

In the traditional notion of state spaces, the "next steps" are part of the "original" state space; all future states are, by definition, within the state space. The specific next state is given by the equations (laws) that determine the evolution of the system, that is, the position it occupies once the system moves from time $t$ to $t + 1$. This is the case also for non-autonomous systems, even if the evolution involves some stochastic component. In this case, the equation determining system evolution can be thought of as having a probabilistic component that is realized at the time in question (e.g., a die is rolled).

**Figure 3**. Adjacent possible. From the actual state, the system can only evolve to adjacent possible states. However, future states of the system are not predictable from the actual state before the system evolves to a given adjacent possible state.

In contrast, by establishing a conceptual bridge between current states and potential future states, the adjacent possible offers a way to model systems that are neither completely random nor entirely predictable, which is a hallmark of many complex biological phenomena. If one accepts Kauffman's argument, some systems can only explore and access the adjacent possible: the set of all possible next steps that are *one step away* from its current state, because in truly adaptive systems the adjacent possible is constantly expanding and evolving, as each step forward opens up new possibilities for the system to explore (Figure 3). This means that the system is constantly discovering new opportunities and adapting to its changing environment, even as it remains constrained by the current state.

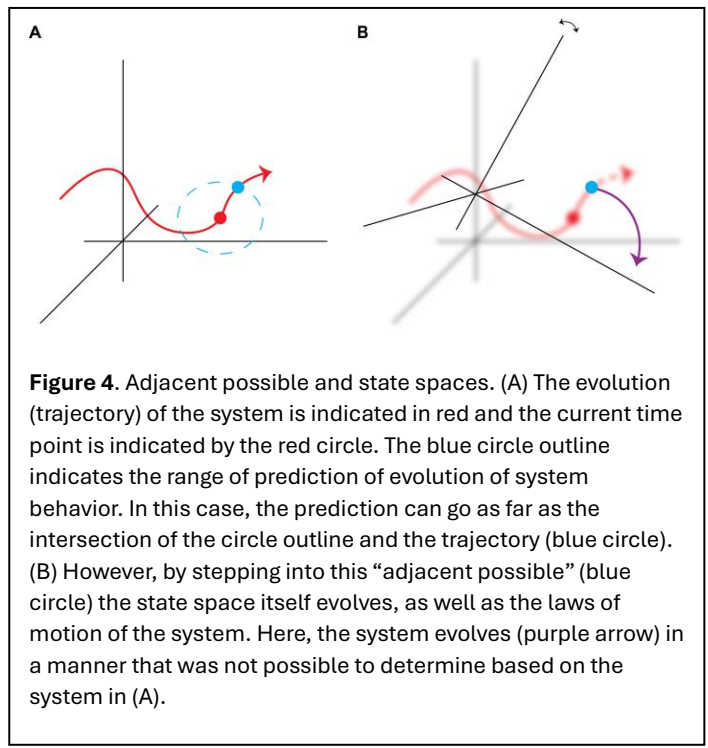

**Figure 4**. Adjacent possible and state spaces. (A) The evolution (trajectory) of the system is indicated in red and the current time point is indicated by the red circle. The blue circle outline indicates the range of prediction of evolution of system behavior. In this case, the prediction can go as far as the intersection of the circle outline and the trajectory (blue circle). (B) However, by stepping into this "adjacent possible" (blue circle) the state space itself evolves, as well as the laws of motion of the system. Here, the system evolves (purple arrow) in a manner that was not possible to determine based on the system in (A).

In terms of state spaces, the implication is these are themselves *dynamic* (Figure 4). Overall, the concept of the adjacent possible suggests that the temporal evolution of complex systems is not a straightforward process, but rather a complex and unpredictable exploration of the possibilities that are available within each system's unique set of constraints and capabilities. In complex systems that follow such radical behavior, one cannot enumerate in advance (what Kauffman calls "*pre-state*") all of the conditions that determine its future. In other words, from actual (present) states, it is not possible to determine future states without stepping into *adjacent possible* states, which *open up possibilities for further system evolution*. In all, the system's evolution is highly, even radically, context dependent. That is to say, how the system evolves is strongly dependent on its history---current states carry with them the history of the system.

### 3.3 Closed worlds of experiments versus natural open worlds

The notion of open-endedness and particularly the *adjacent possible* might strike the reader as very extreme. My goal here is not to advance them in a manner that the reader should accept wholesale. Instead, my objective is to discuss concepts that challenge the notion that we can effectively formalize systems in terms of network models even when they are aided by mathematical formalisms more broadly conceived. Put more positively, they also provide ways to think of new classes of models that consider the problems discussed.

In this section, I provide further arguments aimed at strengthening the notion of open-endedness in complex biological systems by providing a brief discussion of the distinction between *closed vs. open worlds*. For a general discussion of the related notion of small and large worlds (Savage, 1954), see Jaeger (2024, Chapter 5); see also Jaeger et al. (2024)

For concreteness, consider Pavlovian *fear learning*, where an initially neutral stimulus (light) repeatedly paired with an aversive stimulus (shock) acquires affective significance. Once such learning occurs, exposure to the light leads to a response (called the conditioned response) that shares many of the elements generated by an inherently aversive stimulus. However, when such conditioned stimulus (light) no longer predicts the unconditioned stimulus (shock) to which it was paired in the past (i.e., the light is no longer followed by shock), the conditioned stimulus gradually stops eliciting the conditioned response. This process is called *fear extinction*.

In the laboratory, a basic extinction procedure will aim to make the situation as unconfounded by "extra" variables as possible, and the animal is placed in a minimal world that is as clean as possible. In this "closed world", it is possible to describe how the brain-behavior state evolves across time in a relatively well-defined way. For example, with repeated presentations of the light without an accompanying shock, the animal freezes in place less frequently ("freezing" is the term for the behavior of immobility triggered by strongly threating stimuli). Overall, the behavior evolves in predictable ways and most of the action in the brain involves just a few areas. In other words, during the process of extinction, we can delineate a well-defined trajectory through a brain-behavior state space.

However, the situation is rather distinct in the "open world" animals inhabit in nature. Brain-behavior trajectories in the wild will be *variable and unpredictable*. This is because the exact trajectory depends on the animal's history, the state of the world, and the internal state of the animal. In the laboratory, upon detecting the conditioned stimulus, the animal freezes in place---the *only* behavior that is possible. In natural settings, animals can potentially escape impending threats, and diverse and rich behaviors are observed (Evans et al., 2019; Branco and Redgrave, 2020). Multiple behaviors are possible, and they are selected based on multiple interdependent factors. In particular, I have discussed how the neural circuits understood to be involved in fear extinction have been considerably expanded by neuroscientists in the past decades (Pessoa, 2022, chapter 11), as well as the difficulties of trying to delineate a "fear extinction circuit" in the brain (Pessoa, 2023).

Let's briefly discuss possible objections to the notion that relatively more closed and relatively more open worlds paint substantially distinct pictures of brain-behavior relationships. Specifically, although there is evidence for the variability of animal behaviors, they fall into relatively well-defined classes, including species-typical behaviors (Huskisson, 2022). For example, in adult female and male lion interactions, one observes a diverse but finite series of common behavioral interactions. In other words, the objection is that biological systems operate in a sufficiently constrained manner.

This is of course a reasonable reply. Nevertheless, one can only speak of "average behaviors", as in the following situation: consider the exchange between a healthy adult male lion and a female hyena (note that female hyenas are dominant over males of their species). Given a specific condition (ages and sizes of the individuals involved, their distances, potential access to a food source, hunger status, numbers of members of the same species nearby, etc.), one can only say that at a specific moment the hyena is likely to back down and retreat. But individual behaviors are too variable, and "repetitions" of the situation will lead to different outcomes[7]; the hyena could maintain the same distance for longer, approach further or even attack[8].

In sum, the notion of open worlds in biological settings, and in particular in the context of brain and behavior, provide credence to the claim that many biological systems are radically open-ended to allow formalization with standard mathematical frameworks (which would require, at a minimum specification of all potential subtypes of behavior and interactions with the world in advance). This is, admittedly, a strong claim, and it could be argued that strategies employed in complexity science and computational physics could prove successful. As briefly discussed previously, the success of climate models, for one, derives in part from the extremely rich parametrization of the models that overall allow the handling of a large number of interdependent factors. In any event, we know that the success of climate models is partial, and I suggest that complex biological systems pose comparable challenges, and likely greater ones.

## 4 Objections and possible ways forward

---

[7] These considerations naturally connect to questions about determinism. Here, what I mean is the following: given that the exact conditions of the environment, bodies, and nervous systems involved, as well as the ontogenies of the animals involved, will almost-never repeat, there is considerable variability of the precise behaviors generated when conditions are "repeated". In essence, the point is that, in many cases, behaviors are too context and history dependent.

[8] It is worth noting that unseen behaviors have been documented across many species, especially in novel contexts; e.g., bottlenose dolphins have been observed using sponges as tools to protect their rostrums while foraging for fish in the sandy seafloor (Krützen et al., 2005).

The present section discusses a few issues not sufficiently addressed above. Importantly, I will pay particular attention to potential objections to some of the problems discussed above and will discuss potential strategies to address them, at least in part.

### 4.1 Instead of specifying the laws of the system, learn them

The success of physics has relied in great measure on the ability to derive formal principles that were translated into "laws of motion" across many domains of interest. In contrast, in many biological systems, the evolution equations are unknown---or at least remain unknown to this day. What is more, in many cases the system itself evolves dynamically. In any case, it is uncontroversial that the theoretical modeler faces daunting obstacles.

Intriguingly, a potential way around this impasse may be to employ mathematical approaches that, given enough data, learn the dynamics of the system. In other words, although it may prove insurmountable to figure out the "laws of motion" of a biological system, given enough data of the system's behavior, it may be possible to estimate from the data how the system works. Slightly more formally, we can express this situation as learning a function (or set of functions) $f_L$ such that the temporal evolution of the system can be summarizes as $x(t+1) = f_L(x(t))$, where the subscript $L$ denotes that the function was estimated from observing enough data of the system's behavior.

What kinds of system dynamics can be learned in this manner? Indeed, this is a very active area of research, and considerable progress has been made in the past decade (Brunto et al., 2016; Chen et al., 2018; El-Gazzar and van Gerven, 2024). Some techniques have stronger constraints and consider functions $f$ that are relatively fixed. In such cases, the procedure permits estimating parameters that allow the function to reproduce a system's behavior as long as it conforms to the types of behavior captured by $f$. As a trivial example, imagine that $f$ is simply the cosine function and that the periodicity and the starting position of the system are learned from data samples. As long as the system is periodic, this method should be able to reproduce the observed behavior.

Obviously, some systems exhibit considerably more complex behaviors, including strong dependence of the dynamics of the system on its particular state. The fields of control theory and dynamical systems have generated a range of models that come to the aid (Murphy, 2022). Techniques like Hidden Markov Models have the ability to estimate different "states" when enough data are available. With this technique, the system is assumed to transition between a series of states that capture the main modes of operation of the system (Rabiner and Juang, 1986). These models can be extended so that for every state the "laws of motion" of the estate are estimated. In other words, like Hidden Markov Models, they can partition the observed data into a series of states that capture their particularities (e.g., based on video data of an individual at a park, a system may partition the footage in terms of stages of "walking", "running", and "sitting"). In addition, for each state, they can learn additional information that captures the dynamics inherent to that state (e.g., the dynamics of walking, running, and sitting). Such techniques are at times referred to as Switching Linear Dynamical Systems (Ghahramani and Hinton, 2000; Murphy, 2022). In this case, we can describe a system as learning, for a particular system, $x(t+1) = f_{L,s}(x(t))$, where the new subscript $s$ indicates the learning of the dynamics during a specific state. (In principle, the technique allows modeling nonlinear dynamics by approximating nonlinear behavior in terms of a series of linear components.)

Despite considerable progress in the development of such techniques, it is unclear whether they will be able to generalize to large-scale problems[9]. In any case, this work demonstrates the value of data-driven approaches in addressing problems of interest. In principle, these approaches can capture some of the strong context dependence that is observed in biological systems.

In summary, many techniques have been developed that can attempt to estimate the "laws of the system" based on available data. Whereas this general approach is quite powerful, its shortcomings deserve mention. In most cases, although we can envision being able to reproduce and predict system behavior, the models tend to function in a largely "black box" manner, and it is very challenging to interpret the parameters in a manner that aids understanding[10]. To illustrate, the situation would be like learning to predict the evolution of planets around the sun but not be able to distill the knowledge into Kepler's laws, not to mention the reason that these laws work in the first place, namely, Newton's law of universal gravitation. Despite their promise, epistemically, data-driven techniques are frequently on shaky grounds.

*History dependence*

An important scientific goal is to predict the evolution of a system in new situations. Many biological systems exhibit deep history dependence, such that to know $x(t+1)$ one needs to know simultaneously $x(t), x(t-1), x(t-2), \ldots$. In fact, in some situations, the current state could depend on most of the history of the system.

Modeling systems with memory requires different approaches than those used for memoryless ones. The key challenge is to incorporate the system's history into the model in a way that captures the relevant information without becoming computationally intractable. There are several ways to model such systems, including autoregressive models, recurrent neural networks, Markov chains with higher-order memory, and Volterra series (Boyd and Chua, 1985; Murphy, 2022). To illustrate, let's consider the case of a linear autoregressive model, where $x(t)$ depends on multiple time points in the past:

$$x(t) = c_0 + c_1 x(t-1) + c_2 x(t-2) + \cdots + c_p x(t-p).$$

In this case, $x$ at time $t$ depends on $p$ time points in the past, such that each past time point is weighted by a constant $c_i$. This is a simple example, of course, but makes the dependence on past states explicit. As in the cases discussed in the previous section, the parameters $c_i$ are estimated based on data available to the investigator.

---

[9] Among many other difficulties, there are problems of identifiability (large condition numbers) and problems of statistical power.
[10] Briefly, many of the techniques involve learning large numbers of parameters that are used to capture the behavior of the systems in question. However, knowing these parameters does not easily translate to deeper understanding. For example, artificial neural network models might employ hundreds of thousands (and even billions) of parameters, but frequently function largely as black boxes.

In the past decade, recurrent neural networks have become a popular method to estimate future states that depend on (multiple) past states given the development of powerful algorithms capable of determining such time dependencies from observations. A particular class of recurrent neural networks called "reservoir computing" has proven particularly impressive at capturing the dynamics of complex dynamical systems (Pathak et al., 2018; Gauthier et al., 2021). Furthermore, the problem of history dependence has an extensive literature in physics, including non-Markovian processes (Zwanzig, 2001) and non-ergodic systems (Henkel et al., 2008, 2010). More generally, in many biological systems, time dependencies are notoriously difficult to derive from first principles, and accordingly, may necessitate using data-driven methods.

*Coarse graining*

*Adaptive systems* are complex systems that can adjust their behavior and/or structure in response to changes in their environment or internal states. In these systems, the individual components (e.g., species in an ecosystem, people in a social network, or neurons in the brain) collectively determine macro-level properties through local interactions. I discussed in this paper how biological systems of this kind can present considerable challenges to standard network modeling, and even mathematical formalization more generally.

It could be argued, however, that *coarse-graining* considerably ameliorates such challenges. Coarse-graining refers to the process of reducing the level of detail in a system's description by grouping or averaging over microscopic (lower level) details. The key idea is that macro-level properties that emerge from collective properties of the system's components are not arbitrary but, rather, capture important regularities of the system's behavior. These macro properties smooth over (or screen off) the microscopic details; they do not include every small-scale interaction or fluctuation. However, they still capture relevant patterns that can help predict the system's future state, at least locally (i.e., over short time scales or in the vicinity of the current state). Thus, it is possible to provide a simplified, coarse-grained description of a system's dynamics through the collective behavior of its components. Moreover, such description is useful for understanding and predicting the system's behavior, although it does not include every microscopic detail.

To the extent that coarse-graining is successful in investigating complex systems (Sethna, 2021; see also Flack, 2016), the challenges are greatly reduced. However, whereas the approach can be effective in many cases, it does not provide a failproof solution. Indeed, coarse-graining can fail in multiple situations, as briefly outlined next.

(1) *Loss of essential information*. Naturally, if the coarse-graining procedure averages over microscopic details that are crucial for understanding the system's behavior, the resulting simplified description will not capture the relevant dynamics. This can happen when the microscopic details have a significant impact on the system's evolution, such as in chaotic systems where small differences in initial conditions lead to drastically different outcomes (Strogatz, 2018). Additionally, in some cases, like protein folding, it is not that small changes in the initial conditions lead to drastically different outcomes, but rather that the microscopic details themselves are essential for understanding and predicting the folding behavior (Dill and MacCallum, 2012).

(2) *Emergent phenomena*. Some systems exhibit emergent properties that arise from the collective behavior of their components but cannot be directly inferred from the individual components' properties. If the coarse-graining procedure fails to capture these emergent phenomena, the simplified description will miss important aspects of the system's behavior[11].

(3) *Non-equilibrium systems*. Coarse-graining methods often rely on the assumption that the system is in or near equilibrium, namely, that its macroscopic properties do not change significantly over time. When applying coarse-graining to non-equilibrium systems, the reduced set of variables may not capture the system's dynamic behavior accurately; the coarse-grained model might fail to predict long-term trends or emergent properties; and time-dependent effects that are crucial to the system's function might be lost in the simplification. Yet, many complex systems operate far from equilibrium and exhibit non-stationary dynamics, such as gene regulatory networks, immune system dynamics, as well as neuronal circuits (Juarrero, 1999).

(4) *Multi-scale interactions*. When a system has important interactions or feedback loops across multiple spatial and/or temporal scales, coarse-graining at a single scale proves insufficient. In these cases, a multi-scale approach that captures the interplay between different levels will be necessary. In all likelihood, this applies to attempts to understand nervous systems.

(5) *Changing environments*. If the system's environment or external conditions change significantly over time, the coarse-grained description that was valid at one point may no longer be accurate. In such cases, the coarse-graining procedure may need to be adaptive and updated as the system evolves. As an example, consider the challenges of modeling ecosystem dynamics under climate change (Thuiller et al., 2008).

Taken together, while coarse-graining can be quite powerful and aids understanding of complex systems, it can fail when the simplified description does not capture the essential features of the system's dynamics due to the loss of crucial microscopic details, the presence of some types of emergent phenomena, non-equilibrium behavior, multi-scale interactions, and/or changing environments.

## 5 Conclusions

This paper has argued that while network models have been tremendously useful in studying a wide range of phenomena across biological, social, and technological domains, they fall short when applied to complex biological systems, particularly in neuroscience. Here, I have explored several key limitations of traditional network approaches and discussed potential extensions and alternative frameworks that may be better suited to capture the intricate dynamics of adaptive biological systems.

---

[11] Strictly speaking, the point on emergence is directly tied to the preceding one. In addition, some types of emergence are well captured by coarse graining. For a trivial example, consider that temperature in a gas is a macroscopic property that quantifies the average kinetic energy of its constituent particles, rather than a characteristic attributable to any individual particle.

The limitations of basic network models identified include:

1. The need to account for time-varying connections and adaptive topologies.
2. The challenge of representing multi-level systems with cross-level interactions.
3. The difficulty in capturing the radical context dependence exhibited by many biological systems.
4. The inadequacy of fixed state spaces for systems that continuously expand their range of possible states.
5. The challenge to represent the deep history dependence often observed in biological phenomena.

I also discussed how even powerful approaches that take into account time-varying dynamics and context dependence (e.g., along the lines of non-autonomous systems), while powerful, still fall short in fully capturing the complexity of adaptive biological systems. An important reason for this failure relates to the shortcomings of coarse-graining adaptive biological systems.

The implications of these limitations are very substantial, particularly in neuroscience. Our understanding of brain function and behavior may be fundamentally constrained by our current modeling approaches. The brain-body-environment coupling, the impact of neuromodulators on circuit function, and the stark differences between closed experimental setups and open natural environments all point to the need for more flexible, adaptive modeling frameworks.

Moving forward, several promising directions emerge:

1. Developing frameworks that explicitly account for the adaptive nature of biological systems, allowing for dynamic reconfiguration of both nodes and connections.
2. Exploring data-driven approaches that can learn system dynamics without requiring a priori specification of all possible states or interactions.
3. Investigating methods to incorporate deep history dependence and context sensitivity into models.
4. Pursuing multi-scale modeling approaches that can capture interactions across different levels of organization.
5. Embracing the open-ended nature of biological systems, perhaps by drawing inspiration from concepts like Kauffman's "adjacent possible."

While this paper has focused on the limitations of network models, I do not advocate for their wholesale abandonment. Rather, I argue for an approach where network models are extended in the direction of adaptive frameworks. As stated in the Introduction, whether one considers "network science" in a more restrictive or expansive manner will in part determine the extent to which the critiques outlined here apply to this domain of knowledge. In any case, several of the considerations apply to mathematical formalization in general, irrespective of the terminology adopted.

In conclusion, while the path forward is yet to be developed, recognizing the limitations of our current approaches is a necessary first step. Pushing beyond traditional network models and embracing the adaptive, context-dependent nature of biological systems will foster the development of new frameworks that offer deeper insights into the complexities of biological systems generally, and of brain function and behavior in particular. Ultimately, network science needs to develop mathematical and computational frameworks that are more radically dynamic and open ended.

**Acknowledgements**

I thank Manlio De Domenico for incisive feedback on an earlier version of the manuscript. I'm also grateful for feedback by David Barack, as well as by Konrad Kording on the section "Objections and possible ways forward".

**Captions**

Figure 1. Multiple levels in neuroscience. Levels illustrated from molecules to multi-area circuits. The empty circles represent neurotransmitters and filled circles represent receptors. Within an area, different cell populations exist (here based on receptor types).

Figure 2. Context dependence and network nodes. (A) Initial configuration with nodes $i$, $j$, and $k$. (B) In a new context, node $i$ needs to be considered in terms of two subunits, $i_1$ and $i_2$. In this scenario, the network has four nodes. (C) In this context, the original nodes are insufficient to model the situation, and additional elements are needed, including node $l$ and its connections, as well as possibly other nodes and their connections.

Figure 3. Adjacent possible. From the actual state, the system can only evolve to adjacent possible states. However, future states of the system are not predictable from the actual state before the system evolves to a given adjacent possible state.

Figure 4. Adjacent possible and state spaces. (A) The evolution (trajectory) of the system is indicated in red and the current time point is indicated by the red circle. The blue circle outline indicates the range of prediction of evolution of system behavior. In this case, the prediction can go as far as the intersection of the circle outline and the trajectory (blue circle). (B) However, by stepping into this "adjacent possible" (blue circle) the state space itself evolves, as well as the laws of motion of the system. Here, the system evolves (purple arrow) in a manner that was not possible to determine based on the system in (A).